\documentclass[amsfonts,preprint]{iopart}
\usepackage{graphicx}
\begin{document}

\title{Glass-forming liquids: One or more ``order'' parameters?}


\author{Nicholas P. Bailey, Tage Christensen, Bo Jakobsen, Kristine Niss, Niels Boye Olsen, Ulf R. Pedersen, Thomas B. Schr{\o}der, and Jeppe C. Dyre}
\address{DNRF Centre ``Glass and Time,'' IMFUFA, Department of Sciences, Roskilde University, Postbox 260, DK-4000 Roskilde, Denmark}
\date{\today}

\begin{abstract}
We first summarize the classical arguments that the vast majority of glass-forming liquids require more than one ``order'' parameter for their description. Critiques against this conventional wisdom are then presented, and it is argued that the matter deserves to be reconsidered in light of recent experimental developments. Out of the eight basic thermoviscoelastic frequency-dependent response functions, there are generally three independent functions. For stochastic dynamics we show that there are only two independent response functions; for this case it is shown how analytic continuation may be utilized to express the third response functions in terms of two others. Operational criteria are presented for the linear thermoviscoelasticity being described by a single ``order'' parameter, in which case there is just one independent thermoviscoelastic response function. It is shown that a single ``order'' parameter description applies to a good approximation whenever thermal equilibrium fluctuations of fundamental variables like energy and pressure are strongly correlated. Results from computer simulations showing that this is the case for a number of simple glass-forming liquids, as well as a few exceptions, are briefly presented. Finally, we discuss a new conjecture according to which experiments at varying temperature and pressure follow the density-scaling expression for the relaxation time, $\tau=F(\rho^x/T)$ ($\rho$ and $T$ are density and temperature), if and only if the liquid is ``strongly correlating,'' i.e., to a good approximation is described by a single ``order'' parameter.
\end{abstract}

\newcommand{\pf}[2]{{\frac{\partial#2}{\partial#1}}}
\newcommand{\vektor}[1]{\left(\begin{array}{c}#1\end{array}\right)}
\newcommand{\nmatrix}[2]{\left(\begin{array}{#1}#2\end{array}\right)}
\newcommand{\dd}[1]{\delta{#1}}
\newcommand{\half}{\frac 1 2}
\newcommand{\ratio}{\Lambda} 
\newcommand{\na}{{\alpha}} 
\newcommand{\nb}{{\beta}}

\maketitle

\section{Introduction}

The question whether one ``order'' parameter is sufficient for describing glass structure attracted considerable interest among the limited number of glass scientists in the period 1950-1980. The question was thoroughly discussed in particular in the 1970's \cite{ gup76,moy76, roe77,moy78,les80,moy81} leading to clarifications of a number of theoretical questions. Since then, based on experimental evidence the consensus has been that one ``order'' parameter is rarely enough.

The term ``order parameter'' was commonly used in the glass community before the term in the 1960's became commonly known in the physics community where it took on a somewhat different meaning. In connection with critical phenomena and the theory of second order phase transitions, renormalization, etc, ``order parameters'' reflect the relevant Lie group symmetry and determine the relevant part of the free energy within a standard Ginzburg-Landau expansion of the free energy. In order not to confuse the issue it is probably a good idea to change the wording, so below we refer to {\it  ``order'' parameters} or occasionally just {\it parameters}.

The present paper summarizes and extends recent works making the case that the question of how many ``order'' parameters are sufficient deserves to be reconsidered. In Sec. 2 we briefly summarize the classical viewpoint, in Sec. 3 critiques against it are presented, in Sec. 4 the more restricted and well-defined case of linear thermoviscoelasticity is presented, in Sec. 5 thermoviscoelasticity in complete generality is discussed, in Sec. 6 we show that in any stochastic description of the dynamics there are only two independent response functions, Sec. 7 treats the single-parameter case where there is just one independent thermoviscoelastic response function, in Sec. 8 the new concept of a ``dynamic'' Prigogine-Defay ratio, which tests one-parameter-ness by reference to single-frequency thermoviscoelastic measurements, is presented.  Section 9 presents a few computer simulations showing that several systems indeed are well described by only a single parameter, Sec. 10 discusses a recent conjecture stating that the single-parameter liquids are precisely those that obey density scaling for the results of high-pressure experiments. Finally,  Sec. 11 gives a brief summary.

\section{The conventional wisdom: One parameter is seldom enough}

The standard ``order'' parameter theory of glass science was developed by Davies and Jones in the 1950's \cite{dav52,dav53}. This theory idealizes the glass transition and treats it as a genuine phase transition. In the liquid the ``order'' parameters are functions of pressure and temperature, whereas they are frozen in the glass phase. If $\Delta c_p$ is the difference between liquid and glass isobaric specific heat per unit volume at the glass transition temperature $T_g$, $\Delta\kappa_T$ the liquid-glass difference of isothermal compressibilities, and $\Delta\alpha_p$ the liquid-glass difference of isobaric thermal expansion coefficients, the Prigogine-Defay ratio $\Pi$ is defined \cite{dav52,dav53,pri54} by

\begin{equation}\label{pdf}
  \Pi \ =\  \frac{\Delta
  c_p\Delta\kappa_T}{T_g\left(\Delta\alpha_p\right)^2}\,. 
\end{equation}
Within the Davies-Jones framework one can prove \cite{dav52,dav53} that $\Pi\ge 1$, an inequality that has been confirmed in many experiments on quite diverse glass-forming liquids \cite{bra85,sch86,gut95,don01}. If there is just a single ``order'' parameter, one has $\Pi=1$. Although there are glass-forming polymers where $\Pi=1$ within experimental uncertainty \cite{oel77,zol82}, the vast majority, if not all, glass-forming liquids have $\Pi >1$ (typically: $2<\Pi<5$) \cite{moy76}. If simple first-order dynamics are adopted, the case of a single parameter implies an exponential decay towards equilibrium after external disturbances \cite{gup76,moy76,roe77,moy78,les80,moy81,dav52,dav53}. This is rarely observed, a fact that traditionally was seen as a confirmation of the conventional wisdom that more than one parameter is required.

A further classical argument for one parameter not being enough is the well-known fact that glass properties are not uniquely defined by, e.g., the density, as one would expect if there is just one parameter \cite{roe77,sch86}. For instance, one can prepare glasses with same index of refraction, but different electrical conductivity. This point was beautifully illustrated in Kovacs' classical cross-over experiments \cite{sch86,kov63}.

In summary: The observed Prigogine-Defay ratios are almost always significantly larger than unity, relaxations are almost always nonexponential, and glass properties are not just a function of density. This altogether makes a convincing case for there generally being a need for more than one parameter, a conclusion that also appears natural given the complexities of glass-forming liquids and glass structure. Based on this, with few exceptions (e.g., Ref. \cite{nie97,sam99,jah06,sch06}), the matter has not been actively discussed for long time.

\section{Questioning the conventional wisdom}

The first point to be noted is that the question of one or more ``order'' parameters is not really well defined in the classical approach, because the glass transition is not a phase transition. The fact that the glass transition is a dynamic phenomenon -- a gradual falling-out-of-equilibrium that inevitably takes place whenever inherent relaxation times become longer than experimental times -- is well known and well understood. This weakens the classical theory where one regards the glass transition as a freezing-in process taking place at a particular temperature \cite{sim31}.

A related conceptual problem is that the $\Pi$ of Eq. (\ref{pdf}) is not strictly well defined. The changes in specific heat, etc, from liquid to glass are not well defined because of two facts: 1) These changes are found by extrapolating the liquid and glass properties, respectively, to the transition region. The glass transition temperature, however, is not strictly well defined because the glass transition is not a phase transition. 2) The glass phase is not well defined -- and it relaxes continuously -- in principle making any measured property in the glass phase a function of time. Many researchers would argue that, while this is correct in principle, these effects are minor and not sufficiently important to reduce the observed Prigogine-Defay ratios to unity. We take a more purist viewpoint, however, and believe that concepts that are not well defined should be avoided in a scientific description.

Recent experiments monitoring ageing of a glass at temperatures around $T_g$ indicate that in some cases the deviations from equilibrium may be quantified in terms of a single parameter. One example is a study where the characteristics of the dielectric Johari-Goldstein beta loss peak were used to monitor structural relaxations taking place on the alpha-time scale \cite{ols98,dyr03}. To keep things simple only beta loss peak frequency and beta maximum loss were monitored, thus providing two numbers that depend on structure and temperature. For both sorbitol \cite{ols98} and tripropylene glycol \cite{dyr03} it was found that at any given temperature these two numbers correlate linearly. Thus even after a complex thermal history, when returning back to some given temperature, beta loss-peak frequency and loss maximum always lie on a line characterizing that temperature. An example of this is provided in Fig. 1 showing loss-peak frequency and maximum loss for the Johari-Goldstein beta process of tri-propylene glycol during a temperature cycling around $T_g$. If the structure were characterized by more than one order parameter, there is no reason why such a correlation should hold. On the other hand, if structure is characterized by a single parameter, at any given temperature the two quantities must correlate, and for fairly small deviations from equilibrium this correlation would appear approximately linear.

\begin{figure}\begin{center}
\includegraphics[width=8cm]{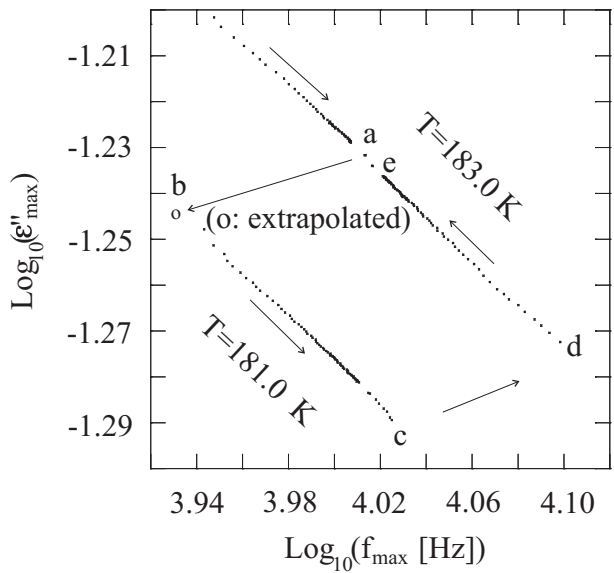}
\caption{Temperature-jump experiment for tripropylene glycol monitored via the beta loss-peak frequency and loss peak maximum \cite{dyr03}. These two numbers depend on structure, and their relaxation monitors structural relaxation (known to take place on the alpha-time scale). Starting at 185.0 K temperature was first lowered to 183.0 K and kept there for 84 h. Then temperature was changed to 181.0 K where it was kept for 140 h. Thereafter temperature was changed back to 183.0 K and kept there for another 140 h. The fact that the beta loss-peak frequency and loss maximum correlate at any given temperature, also after jumping to 181.0 K and back go 183.0 K, indicates that the structure controling the beta relaxation may be described by a single parameter.}
\end{center}\end{figure}

Other dielectric experiments also indicate that a single structural parameter may be sufficient in some cases. Thus studying the shape of the alpha loss peak as quantified by the exponent of the best-fit stretched exponential function, it has been shown for a number of liquids \cite{nga05} that when both temperature and pressure are varied, the shape depends only of the loss-peak frequency. A simple explanation of this would be that there is just a single structural parameter, because if that were the case, this parameter would determine both loss-peak frequency and loss-peak shape and consequently these two quantities would automatically correlate.

Richert and Weinstein from a study of the nonlinear dielectric response on glycerol showed that, although the dielectric and thermal relaxation times vary throughout the liquid, they are locally closely correlated \cite{ric06}. Again, if there is just one parameter determining all properties, one would expect that this parameter may fluctuate in space, but locally determine both dielectric and thermal relaxation time.

More evidence comes from computer simulations. In a model of ortho-terphenyl Mossa and Sciortino \cite{mos04} studied ageing for fairly small temperature steps that were, however, large enough to be well outside the linear regime. The simulations showed that in configuration space the location of the aging system can be traced back to equilibrium states. The authors summarized their findings by stating that for non-linear relaxations close to equilibrium ``a thermodynamic description based on one additional parameter can be provided.''

There seems to be a general understanding in the glass community that there is one ``order'' parameter if and only if the Prigogine-Defay ratio is unity, which happens if and only if relaxations are simple exponentials. This is not correct, however, and not what one finds from reading the classical papers carefully. Goldstein in his 1964 review, for instance, noted that in some situations with several parameters that are mathematically constrained, the Prigogine-Defay ratio may be unity. In this situation ``it is really a matter of taste'' \cite{gol64} whether one prefers to speak of many (constrained) parameters or of a single {\it generally non-exponential} ``order'' parameter. Thus the observation of non-exponential relaxations does not imply that there must be more than one parameter. 

In 2006 Schmelzer and Gutzow revisited the Prigogine-Defay ratio and the question of the number of ``order'' parameters \cite{sch06}. Assuming just a single parameter the dynamics of which follow the classical framework of the thermodynamics of irreversible processes, they showed that the standard Prigogine-Defay ratio obtained by extrapolating from glass and liquid to $T_g$ must be larger than unity. This result seriously questions the prevailing understanding of the glass community referred above, and it emphasizes the need for further work.

The first one should do in reconsidering the ``order'' parameter question is to make sure that the problem is well defined. As mentioned, the standard Prigogine-Defay ratio $\Pi$ of Eq. (\ref{pdf}) is not well defined. As became clear in the 1970's \cite{roe77,moy78,moy81}, it is possible to define a version of $\Pi$ that is well defined. This is done by referring exclusively to properties of the equilibrium viscous liquid phase and its linear responses. In this phase thermodynamic properties are generally frequency dependent, and the high-frequency limits correspond to glassy behavior where structural relaxations do not take place. If $c_p(\omega)$ is the frequency-dependent isobaric specific heat per unit volume \cite{chr07}, etc,  this leads to the following rigorous definition of the Prigogine-Defay ratio for the metastable equilibrium viscous liquid at any temperature $T$:

\begin{equation}\label{prigogine}
  \Pi \ =\  \frac{    \big[ c_p(\omega\rightarrow 0)-  c_p(\omega\rightarrow \infty) \big]\big[  \kappa_T(\omega\rightarrow 0)- 
  \kappa_T(\omega\rightarrow \infty)\big]}   {T\big[\alpha_p(\omega\rightarrow 0) -\alpha_p(\omega\rightarrow \infty)\big]^2}\,. 
\end{equation}

\section{Linear thermoviscoelasticity}

From now on we turn the focus exclusively to the metastable liquid phase with no reference to the glass phase. This limits the discussion compared to what is standard in glass science, but has the advantage of making all concepts rigorously well defined. Linear thermoviscoelasticity deals with the frequency-dependence of thermodynamic properties and their coupling to frequency-dependent mechanical properties. It is understood that, in principle, only infinitesimal perturbations are applied, thus ensuring linearity. In the simplest (isotropic) theory there are two fundamental ``energy bonds,'' a thermal and a mechanical. An energy bond has an ``effort'' variable and a ``displacement'' variable \cite{ost73,mik93,pvc}. The thermal energy bond is characterized by entropy $S$ as the displacement variable and temperature $T$ as the effort, for the mechanical energy bond the displacement variable is the volume $V$ and the effort is the negative pressure, $-p$. The product of the effort and the differential displacement variable gives the energy transferred into the system from its surroundings. Thus the two energy bonds (Fig. 2) simply express the well-known fundamental identity $dE=TdS-pdV$.

\begin{figure}
\begin{center}
\includegraphics[width=8cm]{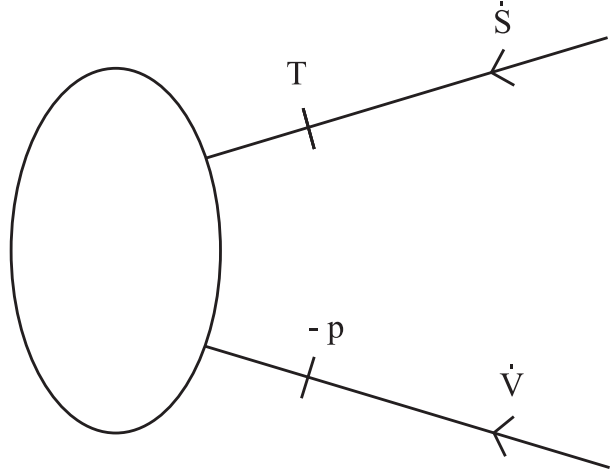}
\caption{The two fundamental energy bonds for a system described by standard thermodynamics. One energy bond is thermal; here the ``effort'' is temperature $T$ and the displacement variable is the entropy $S$; thus if $\dot S\equiv dS/dt$ is the entropy flux into the system, the rate of energy transferred into the system is $T\dot S$. The second energy bond is mechanical; here the effort is negative pressure $-p$ and the displacement variable is the volume $V$; if $\dot V$ is the volume flux into the system, the rate of energy transferred into the system is $-p\dot V$. }
\label{figure_pd}\end{center}
\end{figure}

For infinitesimal perturbations around equilibrium with angular frequency $\omega$, if one imagines controlling the effort variables and measuring displacement changes, and if the usual complex notation is adopted where, e.g., $T(t)=T_0+\delta T(t)$ with $\delta T(t)={\rm Re}[\delta T \exp(i\omega t)]$, linearity is expressed in the following relation where $\delta s$ is entropy change per unit volume and $\delta v$ is relative volume change:

\begin{eqnarray}\label{respfuncTp}
    \vektor{\dd s(\omega)\\ \dd v(\omega)}
    &=&\nmatrix{cc}{
       c_p(\omega)/T & \alpha_p(\omega)\\
       \alpha_p(\omega) & \kappa_T(\omega)
    }\vektor{\dd T(\omega)\\ -\dd p(\omega) }\,.
\end{eqnarray}

The response matrix is sometimes termed the thermal compliance matrix. Its symmetry expresses Onsager reciprocity, reflecting the fundamental fact that there is time reversibility on the microscopic level \cite{kam81,rei98}.

\section{The completely general case: Three independent thermoviscoelastic response functions}

How many independent thermoviscoelastic response functions exist? From the four variables, entropy, temperature, volume and pressure, one may choose any two as ``control'' variables. Usually, one chooses one control variable from each energy bond. There are thus four natural choices of control (``input'') variables, the two remaining are the measured (``output'') variables. For each choice there is one response matrix as in Eq. (\ref{respfuncTp}) \cite{ell07}. These four matrices are all symmetric by Onsager reciprocity, leaving 12 frequency-dependent response functions. These are not independent, however; if one matrix is known, the three others are easily calculated from it by isolating the output variables in question on the left hand sides of two equations. Thus there are only the {\it three} independent response functions, for instance those of Eq. (\ref{respfuncTp}).

The above statement is true in complete generality. Papers by Moynihan and others of the 1970's showed, however, that in the ``order'' parameter description there are really only {\it two} independent response functions \cite{gup76,moy78,ber78} (see also Ref. \cite{mei59}). The formula for calculating the third response function in terms of the two others involves analytic continuation (next section). The situation is analogous to that of the Kramers-Kronig relation which allows one to calculate the imaginary part of a response function in terms of its real part, but only if the latter is known at all frequencies.

\section{The general stochastic case: Two independent thermoviscoelastic response functions}

In this section we summarize the master equation description of viscous liquid dynamics \cite{ell07,sci05,pal82} and show that it implies that there are just two independent response functions. More precisely, it is shown that knowledge of $c_p(\omega)$ and $\alpha_p(\omega)$ at all frequencies allows one to calculate $\kappa_T(\omega)$ except for an overall additive constant giving the high-frequency limit. This is equivalent to the above-mentioned result derived long ago \cite{gup76,roe77,mei59}, but now in a setting that is explicitly consistent with statistical mechanics. 

In a master equation there are states and stochastic transitions between the states. A complete description is provided by the set of probabilities $\{P_n\}$ that the system is in state $n$. This is an ensemble description making it possible to calculate all properties, including the entropy. Following Ref. \cite{ell07} we shall think of each state as an inherent state in the sense of Stillinger and Weber \cite{sti83} (i.e., a potential energy minimum in configuration space), but other state interpretations are also possible. Each state has the vibrational Gibbs free energy $G_n(T,p)$. The ensemble Gibbs free energy that includes the probability dependence is given \cite{ell07,pal82} by

\begin{equation}\label{gibbs}
  G(T,p,\{P_n\})  =  \sum_n P_n   \Big( G_n(T,p)  + k_B T \ln P_n\Big)\,.
\end{equation}
From this one finds the ensemble volume and entropy by the usual thermodynamic relations $V=\partial G/\partial p$ and $S=-\partial G/\partial T$.

The master equation dynamics are given by first order equations in time that are mathematically similar to those of the classical ``order'' parameter description of glass science:

\begin{equation}\label{master}
\dot P_n\  =\ \sum_m W_{nm} P_m  \,.
\end{equation}
The main difference to the ``order'' parameter description is the constraint $\sum_nP_n=1$ and that the present formalism ensures consistency with statistical mechanics. 

The rate matrix $W$ depends on $T$ and $p$ and changes slightly when these variables are perturbed by small time-dependent variations. The same applies for the equilibrium probabilities, $P^{eq}_n\propto\exp[-G_n(T,p)/k_BT]$. According to the principle of detailed balance, which ensures consistency with statistical mechanics as well as time-reversal invariance, the equilibrium probabilities \cite{kam81,rei98} obey

\begin{equation}\label{detbal}
W_{nm}(T,p)P^{eq}_ m(T,p)\,=\, W_{mn}(T,p)P^{eq}_ n(T,p)\,.
\end{equation}
Here temperature and pressure may be arbitrary functions of time. For periodic infinitesimal perturbations from equilibrium the dynamics are perturbed via the transition matrix's dependence on pressure and temperature. The equilibrium probabilities at $p=p_0$ and $T=T_0$ are denoted by $P_n^0$ and the transition matrix at this state point is denoted by $W^0$. 

If $Q_1$ is entropy and $Q_2$ volume, solving the resulting system of equations leads \cite{ell07} to the following expression for the compliance matrix of Eq. (\ref{respfuncTp}) (where $\pf {P_m} {Q_{\na}}$ and $\pf {P_m} {Q_{\nb}}$ are evaluated at $(T_0,p_0)$, the matrix $A(\omega)$ is defined by $A(\omega)\equiv(W^0-i\omega)^{-1}W^0P^0$, $\na,\nb=1,2$): 

\begin{equation}\label{dq}
J_{\na\nb}(\omega)  \ =\ J_{\na\nb}^{\infty}   +\sum_{m,n}\pf {P_n}
{Q_{\na}}  A_{n m}(\omega)  \pf {P_m} {Q_{\nb}} \,.
\end{equation} 
Note that $A(\omega)\rightarrow 0$ for $\omega\rightarrow\infty$; thus $J_{\na\nb}(\omega)\rightarrow J_{\na\nb}^{\infty}$ for $\omega\rightarrow\infty$. Introducing the matrix $Y_{nm}\equiv(P_{n}^0)^{-\half} W_{nm}^0 (P_{m}^0)^\half$, the detailed balance requirement Eq. (\ref{detbal}) implies that $Y$ is symmetric. In terms of $Y$, the matrix $A(\omega)$ is given \cite{ell07} by $A(\omega)=RY(Y-i\omega)^{-1}R$ where $R_{nm} = (P_{n}^0)^{\half} \delta_{nm}$. Thus for the relaxing part of the compliance matrix $\Delta J\equiv J-J^\infty$, if $\partial Q_\alpha$ is the vector whose $n$'th component is $\partial Q_\alpha/\partial P_n(T_0,p_0)$, one has \cite{ell07}

\begin{equation}\label{relaxJ}
\Delta J_{\alpha\beta}(\omega)\,=\,
\left\langle R\,\partial Q_\alpha\left| \frac{Y}{Y-i\omega}\right| R\,\partial Q_\beta\right\rangle\,.
\end{equation}
Adopting the standard ``ergodicity'' assumption that all states are connected by some path of intermediate states, the matrix $Y$ has a one-dimensional eigenspace corresponding to the eigenvalue zero whereas all other eigenvalues are negative \cite{kam81,rei98}. If the eigenvectors of $Y$ corresponding to all the negative eigenvalues are denoted by $|\psi_j\rangle$ with corresponding eigenvalue $-1/\tau_j$, Eq. (\ref{relaxJ}) implies 

\begin{equation}\label{relaxJ2}
\Delta J_{\alpha\beta}(\omega)\,=\,
\sum_j \langle R\,\partial Q_\alpha\ | \psi_j\rangle \langle \psi_j | R\,\partial Q_\beta\rangle \frac{-1/\tau_j}{-1/\tau_j-i\omega}\,.
\end{equation}
Since $(-1/\tau_j)/(-1/\tau_j-i\omega)=1/(1+i\omega\tau_j)$, changing to a continuous notation Eq. (\ref{relaxJ2}) becomes

\begin{equation}\label{relaxJ3}
\Delta J_{\alpha\beta}(\omega)\,=\,
\int_0^\infty \frac{g_\alpha(\tau)g_\beta(\tau)}{1+i\omega\tau}d\tau\,,
\end{equation}
where the functions $g_\alpha(\tau)$ are real, but not necessarily positive. 

By reference to the theory of analytic functions we show below that not all three functions of the compliance matrix are independent. This is intuitively obvious already from the fact that the {\it three} compliance functions are determined by the {\it two} functions $g_1(\tau)$ and $g_2(\tau)$. More precisely the argument goes as follows. The three compliance functions $\Delta J_{\alpha\beta}(\omega)$ are analytic in the complex positive half plane, i.e., where ${\rm Re}(\omega)>0$. Knowledge of such a function at all real, positive frequencies by analytic continuation uniquely determines the function in the complex plane. Equation (\ref{relaxJ3}) shows that there is a branch cut along the positive imaginary frequency axis. Given that $\Delta J_{\na\nb}(\omega)\rightarrow 0$ for $\omega\rightarrow\infty$, the pole distribution on the branch cut uniquely determines the compliance function. More specifically, Eq. (\ref{relaxJ3}) implies that 

\begin{equation}\label{relaxJ4}
\Delta J_{22}(\omega)\,=\,
\int_0^\infty 
\lim_{\omega'\rightarrow i/\tau}\left\{(1+i\omega'\tau)\frac{\Delta J_{12}^2(\omega')}{\Delta J_{11}(\omega')}\right\}
\frac{1}{1+i\omega\tau}d\tau\,.
\end{equation}
Thus knowledge of $\Delta c_p(\omega)$ and $\Delta\alpha_p(\omega)$ at all real, positive frequencies implies knowledge of $\Delta \kappa_T(\omega)$. Similarly, knowledge of $\Delta \kappa_T(\omega)$ and $\Delta\alpha_p(\omega)$ at all real, positive frequencies implies knowledge of $\Delta c_p(\omega)$.

\section{The single-parameter case: One independent thermoviscoelastic response function}

The compliance matrix $J_{\alpha\beta}(\omega)$ reflects both the relaxing responses (completely characterized by $\Delta J(\omega)$) and the instantaneous responses given by the high-frequency limits. Switching to the time domain, if the relaxing responses of the two energy bonds are always proportional, i.e., controlled by a common variable $\delta\varepsilon(t)$, the entropy and volume responses per unit volume are given by expressions of the form

\begin{eqnarray}\label{order}
 \delta  s(t) &=& \gamma_1 \delta\varepsilon(t) +J_{11}^\infty \delta T(t)   -  J_{12}^\infty  \delta p(t) \nonumber\\ 
\delta v(t)  &=&  \gamma_2\, \delta\varepsilon(t) + J_{21}^\infty \delta T(t)  -J_{22}^\infty \delta p(t)  \,. 
\end{eqnarray} 
We refer to this situation as that of a single ``order'' parameter \cite{ell07} and proceed to show following Ref. \cite{ell07} that in this case there is basically just one compliance function. Note that no reference is made to the properties of the glassy state.

For periodically varying fields Eq. (\ref{order}) implies

\begin{eqnarray}\label{43}
 \delta  s(\omega) &=&\gamma_1  \delta\varepsilon(\omega) +J_{11}^\infty \delta T(\omega) -J_{12}^\infty  \delta p(\omega)\nonumber\\ 
 \delta  v(\omega) &=&\gamma_2  \delta\varepsilon(\omega) +J_{21}^\infty \delta T(\omega) -J_{22}^\infty  \delta p(\omega)
\,. 
\end{eqnarray}
The $\varepsilon$-parameter may be expanded to first order as follows:

\begin{equation}\label{42}
\delta\varepsilon(\omega)\,=\,
\Lambda_1(\omega) \delta T(\omega)-\Lambda_2(\omega)  \delta p(\omega)\,.
\end{equation}
Substituting Eq. (\ref{42}) into Eq. (\ref{43}) and using the symmetry of the compliance matrix leads to the identity $\gamma_1\Lambda_2(\omega)+J_{12}^\infty=\gamma_2\Lambda_1(\omega)+J_{21}^\infty$. For the imaginary parts this implies

\begin{equation}\label{44}
\frac{\Lambda_1''(\omega)}{\gamma_1}\,=\,
\frac{\Lambda_2''(\omega)}{\gamma_2}\,.
\end{equation}
When two analytical functions both with branch cuts on the positive imaginary axis of the complex $\omega$-plane have same imaginary part, they are identical except for an overall additive constant. The latter is zero, because the fact that the two functions give the {\it relaxing} part of the responses implies that they both go to zero for $\omega\rightarrow\infty$. Thus $\Lambda_1(\omega)\propto\Lambda_2(\omega)$. By considering the constant pressure and constant temperature cases it now follows easily from Eqs. (\ref{43}) and (\ref{42}) that $\Delta J_{11}(\omega)\propto\Delta J_{12}(\omega)\propto\Delta J_{22}(\omega)$, or:

\begin{equation}\label{prediction}
\Delta c_p(\omega) 
\,\propto\,\Delta\alpha_p(\omega)
\,\propto\,\Delta\kappa_T(\omega)\,.
\end{equation}

In conclusion, in the case of a single ``order'' parameter (Eq. (\ref{order}) there is basically just one independent thermoviscoelastic response function, i.e., knowledge of one of them implies knowledge of the two others except for the overall additive constant giving their high-frequency limits.

\section{``Dynamic'' Prigogine-Defay ratio: A single-parameter test}

In principle, in order to test experimentally whether or not a single ``order'' parameter suffices, one measures the three response functions of the compliance matrix to test whether the relaxing parts are proportional (Eq. (\ref{prediction})). This, however, requires wide-frequency measurements of the thermoviscoelastic response functions, and there are yet no measurements of all three thermoviscoelastic response functions on a glass-forming liquid. (Even the isobaric frequency-dependent specific heat $c_p(\omega)$ has not yet been measured reliably \cite{chr07}. The problem is that, because frozen-in stresses relax on the same time scale that the enthalpy relaxes; establishing truly isobaric conditions is difficult and in most experimental setups the stress tensor is not diagonal.)

Even when methods have been developed for measuring the compliance matrix of Eq. (\ref{respfuncTp}), one may still expect that initial measurements cover only a rather limited dynamic range. This leads to the question: Is it still possible to test the single-parameter conjecture Eq. (\ref{order})? This question was discussed in a recent publication \cite{ell07} where it was shown that, in fact, measurements at one single frequency are enough to test the single-parameter conjecture. Of course, one can never {\it prove} that a single-parameter description is correct in an absolute sense -- it is all a matter of investigating {\it how good} such a description is. In the above-mentioned recent paper \cite{ell07} it was shown that a ``dynamic'' Prigogine-Defay ratio $\ratio_{Tp}(\omega)\ge 1$ exists with the property that, if this quantity is unity at one frequency, it is unity at all frequencies -- which happens if and only if a single-parameter description applies. The dynamic Prigogine-Defay ratio is given by the imaginary parts of the three thermoviscoelastic response functions \cite{ell07} as follows: 

\begin{equation}\label{tp}   \ratio_{Tp}(\omega)\ =\ \frac{c_p''(\omega)\kappa_T''(\omega)}
  {T_0(\alpha_p''(\omega))^{2}}\,.
\end{equation}
In order to minimize uncertainties measurements should preferably be taken at a frequency around the alpha loss-peak frequency, because only here the imaginary parts are significantly different from zero. We expect that if $\ratio_{Tp}(\omega)$ is close to unity in the main relaxation region (e.g., below $1.1$), a single-parameter description applies to a good approximation.

\section{Results from computer simulations}

Recently thermal equilibrium fluctuations were studied in computer simulations of various liquids \cite{ped06,ped07}. In many cases it was found that in constant temperature and volume simulations (the so-called $NVT$ ensemble) pressure and energy fluctuations correlate strongly. More accurately, this applies for the {\it configurational} parts of pressure and energy, the ``virial'' and the potential energy. (The kinetic parts of pressure and energy -- the ideal gas pressure at the given density and temperature, and the kinetic energy -- trivially correlate 100\%, but with a different proportionality constant.) As an example, Fig. 3 shows the thermal fluctuations of virial and potential energy for a standard Lennard-Jones liquid.

\begin{figure}\begin{center}
\includegraphics[width=8cm]{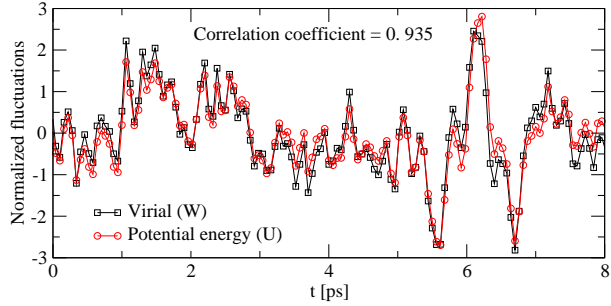}
\caption{Thermal equilibrium fluctuations of potential energy and viral (the configurational part of pressure) for a standard Lennard-Jones liquid \cite{ped07}. The fact that these two quantities correlate strongly shows that, as regards the configurational degrees of freedom, a single-parameter description is quite good for the thermoviscoelastic behavior. For highly viscous liquids the time scale separation between the slow configurational degrees of freedom and the remaining implies that these correlations (that we have also seen, e.g., in simulations of the highly viscous Kob-Andersen binary Lennard-Jones mixture) implies that the three thermoviscoelastic response functions are basically identical.}
\end{center}
\end{figure}

\begin{figure}\begin{center}
\includegraphics[width=8cm]{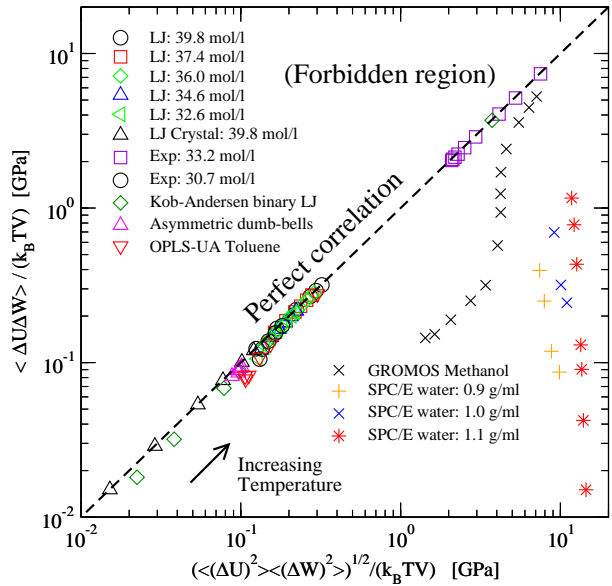}
\caption{Correlation coefficient between virial (volume times the configurational part of the pressure) and potential energy thermal equilibrium fluctuations for a number of liquids evaluated by computer simulations \cite{ped07}. The liquids represented are: LJ: Standard Lennard-Jones, Exp: Monatomic liquid with exponential repulsive forces, Dumb-bell: A molecule model of two atoms of unlike size, BLJ: The Kob-Andersen binary Lennard-Jones liquid, Methanol, and SPC/E Water. The last two are hydrogen bonding and do not show significant correlations; the other liquids do. It has been argued that virial and potential energy give the slowly fluctuating parts of the pressure and energy \cite{ped07}; thus whenever the former quantities correlate strongly, to a good approximation the liquid may be regarded as described by a single parameter.}
\end{center}
\end{figure}

Liquids for which these quantities correlate strongly in their fluctuations are well described by a single order parameter \cite{ell07}. Intuitively this may be understood by reference to Eq. (\ref{order}) considered without perturbations ($\delta T(t)=\delta p(t)=0$) which, if assumed to describe also the {\it fluctuations}, shows that entropy and volume fluctuations are 100\% correlated. Thus one expects that the dynamic Prigogine-Defay ratio is close to unity for such ``strongly correlating liquids.'' Figure 4 shows that this is the case for the Lennard-Jones liquid as well as for a number of other glass-forming liquids. Water and methanol are interesting exceptions that do not show strong correlations between virial and potential energy fluctuations (Fig. 5); thus for these two hydrogen-bonding liquids a single-parameter description does not apply.

\begin{figure}\begin{center}
\includegraphics[width=8cm]{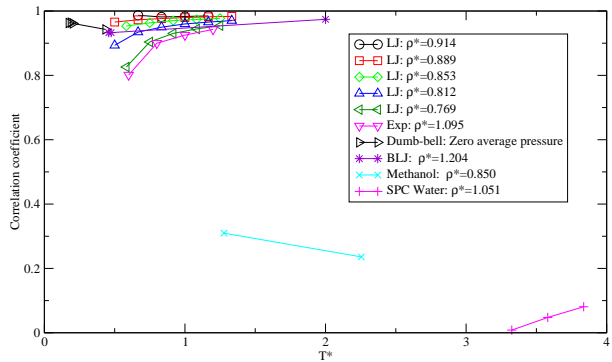}
\caption{Correlation coefficients for a number of glass-forming liquids between virial and potential energy thermal equilibrium fluctuations as function of temperature (in reduced units). The liquids represented are: LJ: Standard Lennard-Jones, Exp: Monatomic liquid with exponential repulsive forces, Dumb-bell: A molecule model of two atoms of unlike size, BLJ: The Kob-Andersen binary Lennard-Jones liquid, Methanol, and SPC/E Water. The figure shows the same systems as those of Fig. 4 studied by computer simulations. The two hydrogen-bonding liquids, water and methanol, show poor correlation, the remaining systems are all strongly correlating. For the latter the correlation even increases as temperature rises; this is because at high temperature the particles approach each other more in collissions than at low temperatures and the inverse power law description of the works better the closer the particles are.}
\end{center}
\end{figure}

\section{A conjecture: {\it Strongly correlating liquids obey density scaling and vice versa}}

The last five years large amounts of data on the behavior of glass-forming liquids under pressure have been published. The motivation is that by not just varying temperature, but pressure as well, much more information may be learned about these systems. Generally, the liquid relaxation time $\tau$, which is basically the Maxwell relaxation time or the alpha loss peak frequency, depends strongly on both temperature and pressure, increasing with lowering temperature or raised pressure. This is not surprising. A new and significant finding \cite{alb04,cas04,rol05}, however, is that if $\rho$ is the density, many liquids obey ``thermodynamic'' or ``density'' scaling, i.e., the function $\tau(T,p)$ may be written

\begin{equation}\label{scaling}
\tau\,=\,
F\left(\frac{\rho^x}{T}\right)\,.
\end{equation}
Both the function $F$ and the exponent $x$ depend on the liquid in question. This expression has mainly been tested on glass-forming molecular liquids, the systems that are most easily accessible. For hydrogen-bonding liquids like glycerol or sorbitol the $x$'s initially reported were anomalously small \cite{rol05}, but it now appears that the reason is that density scaling does not work very well for hydrogen-bonding liquids \cite{grz06}.

Recently, Coslovich and Roland presented computer simulations of binary Lennard-Jones type systems where, however, the exponent of the repulsive term of the potential was varied, taking the values 8, 12, 24, and 36 \cite{cos07}. Such system may be cooled to low temperatures where the viscosity is very large, without crystallizing. Their simulation results obey the density scaling expression Eq. (\ref{scaling}), which by itself is an interesting finding. Even more interesting is the fact that the exponent $x$ may be related to the effective exponent describing the approximate power law of the potential. For the standard binary Lennard-Jones case, for instance, this exponent is not 12 as naively expected, but a number close to 18 depending on the precise choice of fitting criteria \cite{ped07}. 

In Ref. \cite{ped07} some of the present authors previously found that there are strong energy-pressure correlations whenever the repulsive part of the interaction is well described by an inverse power law. Since this seems also to be the criterion for a liquid obeying density scaling (Eq. (\ref{scaling})), an obvious conjecture is \cite{ped07} that: {\it A glass-forming liquid is strongly correlating if and only if it obeys density scaling.} Two liquids that in computer simulations were not strongly correlating are water and methanol \cite{ped07}, and we surmise that hydrogen-bonding liquids generally are not strongly correlating. The argument is that the existence of ``competing interactions'' (van der Waals forces as well as the directional hydrogen bonds) destroy significant correlations, implying that hydrogen-bonding liquids are not well described by a single ``order'' parameter. This is consistent with the finding that hydrogen-bonding liquids do not obey density scaling.

If this conjecture is correct, by virtue of their simplicity the class of strongly correlating liquids provides an obvious starting point for theories for viscous liquids and glass formation. It would be obvious to further conjecture that also covalently bonding liquids are not strongly correlating, again due to the directional nature of the bonds. Many of these systems have fairly low fragility. Low-fragility liquids are traditionally thought to be simple (e.g., have to almost exponential relaxations if the liquid is almost Arrhenius). We here conjecture almost the opposite, namely that many high-fragility liquids in a certain sense are simpler than many low-fragility liquids.

\section{Summary and final remarks}

We have argued that the old discussion of one or more ``order'' parameters deserves to be revitalized. There are indications that at least some glass formers may be well described by a single order parameter as regards their linear thermoviscoelasticity. It is important to emphasize that no claim is made that the molecular structure is completely characterized by a single number. We now have an experimentally useful criterion for whether or not a single-parameter description is accurate. Computer simulations confirm that some model liquids are well described by a single parameter; these liquids are referred to as ``strongly correlating.'' Since hydrogen-bonding liquids do not show these correlations, we expect that liquids with directional bonding are not well described by a single parameter, whereas van der Waals bonded liquids are. Thus we conjecture that for van der Waals liquids the relaxing parts of the three thermoviscoelastic response functions of Eq. (\ref{respfuncTp}) are all proportional, whereas for hydrogen-bonding liquids this is conjectured not to be the case. This prediction can be tested once methods have been developed to measure the full thermoviscoelastic compliance matrix.

\section{Acknowledgments}
This work was supported by a grant from the Danish National Research Foundation (DNRF) for funding the centre for viscous liquid dynamics ``Glass and Time.''

\end{document}